\documentclass[a4paper]{jpconf}
\usepackage{graphicx}
\usepackage{amsmath}
\usepackage[utf8]{inputenc}
\usepackage{cite}
\begin{document}
\begin{flushright} LTH 1110 \end{flushright}

\title{Towards NNLO accuracy for $\varepsilon'/\varepsilon$}

\author{M Cerd\`a-Sevilla${}^1$, M Gorbahn${}^1$, S J{\"a}ger${}^2$ and A Kokulu${}^1$}

\address{ \textit{${}^1$ Department of Mathematical Sciences, University of Liverpool,
Liverpool L69 7ZL, UK}}
\vspace*{2mm}
\address{ \textit{${}^2$ Department of Physics and Astronomy, University of Sussex,
    Brighton BN1 9QH, UK}}

\ead{Maria.Cerda-Sevilla@liverpool.ac.uk, martin.gorbahn@liverpool.ac.uk, S.Jaeger@sussex.ac.uk, Ahmet.Kokulu@liverpool.ac.uk}

\begin{abstract}
The quantity $\varepsilon'/\varepsilon$ measures direct CP violation
in Kaon decays. Recent SM predictions show a $2.9\sigma$ tension with data,
with the theoretical uncertainty dominating.
As rapid progress on the lattice is bringing nonperturbative
long-distance effects under control, 
%
a more precise knowledge of short-distance contributions 
is needed. We describe the first NNLO results for
$\varepsilon'/\varepsilon$ and discuss future prospects, as well as
issues of scheme dependence and the separation of perturbative and
nonperturbative effects.
Finally we also comment on the solution of the renormalisation-group evolution in one of the talks at this conference and present the correct solution.
\end{abstract}

\section{Introduction}
CP violation (CPV) is one of the most fascinating phenomena of high
energy physics and is a natural place to search for physics beyond
the Standard Model (SM).
For instance, the SM is unable to account for the observed matter-antimatter
asymmetry in the universe, which, in thermal baryogenesis scenarios,
requires CPV.
New sources of CPV, however, generally modify the SM predictions for
CPV in flavour-violating decays. In the case
of CPV in $K_L \rightarrow \pi\pi$, the SM predicts
a particularly strong suppression due to
mass and CKM hierarchies, which imply powerful
Glashow-Iliopoulos-Maiani (GIM) cancellations.
This mechanism need not apply to models of new physics.
Therefore kaons are particularly promising for shedding light on
physics at a more fundamental level.
High-precision CP-violating kaon observables offer the
exciting possibility of establishing the presence of new physics (NP). 

While CPV in $K_L$ decays was discovered 50 years ago, and
direct CPV---quantified by the parameter $\varepsilon'$---has
been firmly established for more than a decade, 
the theoretical treatment of the latter suffered from 
nonperturbative uncertainties which until very recently
could not be computed in a controlled approximation.
While short-distance contributions from scales $\sim m_c$ and above
have been known to next-to-leading-order in perturbation theory since
the mid-1990s
\cite{BJLW,BJLW2,BJL,CFMR1,BJL2,CFMR}, the long-distance corrections,
usually represented as $K \to \pi\pi$ matrix elements of local
four-quark operators, have been for the first time
determined with controlled systematics only in 2015
\cite{BBCFG,BBBCF}. This achievement, when combined with the
short-distance calculations, opens the possibility of a
precision theory prediction of $\varepsilon'$. 
A state-of-the-art analysis at NLO within the SM \cite{Analysis} gives
\begin{eqnarray}
\left(\frac{\varepsilon'}{\varepsilon}\right)_{\text{SM}}=(1.9\pm 4.5)\times 10^{-4}.
\end{eqnarray}
(The complex phase is very close to zero.)
This is to be contrasted with the current experimental world
average based on measurements at NA48 \cite{NA48} and KTeV
\cite{KTeV1,KTeV2} of
\begin{eqnarray}
{\rm Re}\left(\frac{\varepsilon'}{\varepsilon}\right)_{\text{exp}}=
     (16.6\pm 2.3)\times 10^{-4} .
\end{eqnarray}
The $2.9\sigma$ difference between experiment and theory
could have one of several origins, the most exciting among
them being a possible contribution
from new particles beyond the SM.  Clearly, a refined SM prediction is
necessary. While at the moment the error budget \cite{Analysis} is
dominated by the uncertainty on a single non-perturbative matrix
element, this error is expected to shrink in the near future with
further progress in lattice QCD. The next most important source of
uncertainty at present is due to missing higher-order perturbative
contributions, making NNLO accuracy imperative. In this proceedings,
we preview (partial) results of an ongoing NNLO calculation
\cite{MMSA}, which roughly cut the (total)
perturbative error in half. We further describe a renormalisation-group-invariant
formalism for separating perturbative and non-perturbative effects and
briefly discuss future perspectives. Finally, we comment on
and correct the expressions for the mixed QCD-QED
renormalisation-group evolution \cite{Kitahara:2016nld} that
were presented in another talk at
this conference.

\section{$\Delta S=1$ Effective Hamiltonian}
The key tool to disentangle the physics of the different
scales ($\Lambda_{\rm QCD}, m_c, m_b, M_W$ and possibly $M_{\rm NP}$) is
the weak effective $\Delta S=1$ Hamiltonian
\cite{BJLW,BJLW2,BJL,CFMR1,BJL2,CFMR}. It takes
the general form
\begin{equation}
  {\cal H}_{\rm eff} = \sum C_i(\mu) Q_i(\mu),
\end{equation}
where the process-independent Wilson coefficients $C_i(\mu)$ contain
the physics of scales (virtualities) above $\mu$ ($\mu^2$), while the
degrees of freedom below $\mu$ are encoded in process-dependent
hadronic matrix elements of local operators $Q_i$, in the present case
$\langle \pi \pi | Q_i(\mu) | K \rangle$. These hadronic matrix elements
are non-perturbative, and the lack of systematic predictions for them has precluded
controlled predictions of $\varepsilon'$ for a long time. This bottleneck has
begun to be removed by recent developments in lattice QCD, in particular
by the RBC and UKQCD collaborations \cite{BBCFG,BBBCF}.

Several choices of the scale $\mu$ are possible. So far, the state of the art
\cite{Analysis} consisted of evaluating the Wilson coefficients in
a three-flavour ($n_f=3$) theory, obtained by integrating out the weak scale,
as well as the bottom and charm quarks. Hadronic matrix elements with
fully quantified systematic errors are available in the $n_f=3$  theory from
RBC and UKQCD \cite{BBCFG,BBBCF} and agree well with earlier calculations in realistic
models \cite{BBG,Bardeen:1986vz,BGB}. In the future, nonperturbative
calculations with dynamical
charm quark will likely become available, allowing to work based on a
weak Hamiltonian in a theory with $n_f=4$ quark flavours.
It is convenient to split
the Wilson coefficients as $C_i = \frac{4 G_F}{\sqrt{2}} V_{ud} V_{us}^* ( z_i + \tau y_i)$, where $z_i$ and $y_i$ are real and
$\tau = - (V_{td} V_{ts}^*)/(V_{ud} V_{us}^*)$ is the phase-convention-independent
CP-violating CKM combination governing CPV in $K$ decays. For
$n_f=4$ (dynamical charm quark), we then have\footnote{Using the projection operators $P_L$ and $P_R$ instead of the traditional $V\pm A\times V\pm A$ introduces a factor of 4 in the overall normalisation of the Hamiltonian.
}
\begin{eqnarray}
  \mathcal{H}_{\text{eff}}=\frac{4G_F}{\sqrt{2}} V_{ud} V_{us}^*\left(\sum_{i=1}^{2}z_i(\mu)[Q_i^u - Q_i^c]+ \tau\left[\sum_{i=1}^{2} z_i(\mu) Q_i^c+\sum_{i=3..10,7\gamma,8g} y_i(\mu) Q_i\right]\right) ,
\label{eq:Hf}
\end{eqnarray}
where we have made use of the fact that   $z_{1,2}^c = y_{1,2}^c = -
z_{1,2}$, while GIM cancellations ensure $z_{3 \dots 10} = 0$.

If the charm quark is integrated out ($n_f=3$), we have instead
\begin{eqnarray}
  \mathcal{H}_{\text{eff}}=\frac{4G_F}{\sqrt{2}} V_{ud} V_{us}^*\left(\sum_{i=1}^{2}z_i(\mu) Q_i^u+\sum_{i=3\dots 10,7\gamma,8g} z_i(\mu) Q_i
  +\sum_{i=3\dots 10,7\gamma,8g}\tau y_i(\mu) Q_i\right).
\end{eqnarray}
The operators $Q_{1,2}^c$ involving the charm quark have disappeared.
The hard GIM breaking in the $n_f=3$ theory now entails nonzero
values also for the ``real parts'' $z_{3 \dots 10,7\gamma,8g}$
of $C_{3 \dots 10,7\gamma,8g}$.

The Wilson coefficients $C_i(\mu)$, or equivalently
$z_i(\mu)$ and $y_i(\mu)$, are determined using
renormalisation-group-improved perturbation theory. Schematically,
\begin{equation}    \label{eq:match}
 C_i(\mu_3) = \left[ U(\mu_3, \mu_c) M^c(\mu_c) U(\mu_c, \mu_b) M^b(\mu_b) U(\mu_b, \mu_W) \right]_{ij} C_j(\mu_W) ,
\end{equation}
which is easily decomposed into expressions for $z_i$ and $y_i$.
The evolution operators $U(\mu_1, \mu_2)$ resum large logarithms $\ln\mu_2/\mu_1$,
while the matching matrices $M(\mu)$ connect the Wilson coefficients
across a flavour threshold, in particular:
\begin{equation}
  C_i^{(n_f = 3)}(\mu_c) = M_{ij}^c(\mu_c) C_j^{(n_f=4)}(\mu_c) .
\end{equation}
The operators can be classified as current-current ($Q^u_{1,2}$, $Q_{1,2}^c$),
QCD penguin ($i=3 \dots 6, 8g$), and electroweak
penguin ($i=7 \dots 10,7\gamma$). 
At present, all ingredients in (\ref{eq:match}) are known to NLO accuracy
(with the exception of the dipole operators $Q_{8g}$ and $Q_{7\gamma}$),
and some of them to NNLO accuracy. The work described here completes the
QCD-penguin part to NNLO, which removes about half of the current perturbative
uncertainty on $\varepsilon'$.

\section{NNLO matching for the QCD penguins}
To obtain the QCD-penguin Wilson coefficients to NNLO accuracy,
we require the submatrices of $M^c$ and $M^b$ for $Q_{1 \dots 6, 8g}$
and $Q_{1,2}^c$.
We leave aside the dipole
operator for the time being, which is formally suppressed by
$m_s/\Lambda_{\rm QCD}$.
The new ingredients are then related to the removal of the operators $Q_1^c$
and $Q_2^c$ at a scale $\mu_c \sim m_c$,
although we have recalculated the remaining,
known \cite{BG} elements of the matching matrices $M^c$, $M^b$. The
anomalous dimension matrices are known from \cite{MH}.

To determine these threshold corrections,
we match Green's functions with operator insertions
in the $n_f$- and $n_f+1$-flavour theories at the matching scale
$\mu_i=\mathcal{O}(m_i)$ ($m_i=m_c, m_b$).

Novel effects at NNLO include
current-current operator inertions getting contributions from virtual
charm/bottom-quarks and the strong coupling constant
being discontinuous at the threshold scale.
In our calculation we use dimensional regularization and
avoid the appearance of traces over $\gamma_5$ by employing the so-called
``modern basis" \cite{CMM,Bobeth:2003at};
however this requires
extra steps to connect with the ``traditional'' basis employed in
the existing literature on $\varepsilon'$, and in the lattice-QCD
calculations, which is nontrivial at NNLO.
We expand in the external momenta (as appropriate for
a matching onto dimension-six operators) and set the masses of
the $n_f$ light quarks to zero. After renormalisation, the $n_f+1$-flavour result
still contains infrared divergences in the form of poles in
$\epsilon = (4 - d)/2$, which have to be reproduced in the $n_f$-flavour  theory.
As the Green's functions in the $n_f$-flavour theory contain only massless
tadpole loop diagrams
(after expansion in the external momenta), they
are given entirely in terms of the ultraviolet
counterterms in the $n_f$-flavour theory, which are related to known
anomalous
dimensions. The cancellation of divergences
constitutes an important check of our calculation. In the end,
the matching results in finite threshold corrections for the Wilson coefficients
\begin{eqnarray}
C_i^{(n_f)}(\mu)=\left[ M_{n_f,n_f+1}(\mu) \right]_{ij} C_j^{(n_f+1)}(\mu).
\end{eqnarray} 

\section{Scale and scheme independence}
Physical observables cannot depend on the dimensional renormalisation
scale $\mu$. This means that the $\mu$-dependence
has to cancel between the Wilson
coefficients $C_i(\mu)$ and matrix elements $\langle Q_i(\mu) \rangle$
order by order in perturbation theory.
However, in the
present case the matrix elements are found from non-perturbative calculations
and their explicit $\mu$-dependence is not known. Instead, we can verify
explicitly
that our Wilson coefficients $C_i(\mu)$ satisfy the NNLO renormalisation-group
equations up to ${\cal O}(\alpha_s^3)$ remainder terms, which they do. However,
this still does not allow to give an estimate of the impact of
missing higher-order perturbative corrections, part of which are
contained in the published lattice-QCD matrix elements and
inseparable from the non-perturbative uncertainties from dynamics at
scales $\sim \Lambda_{\rm QCD}$.

To allow for a full separation of scales, we note that the evolution
operator $U(\mu_1, \mu_2)$ in fact factorizes,
\begin{equation}
  \label{eq:uevolve}
  U(\mu_1, \mu_2) = J(\mu_1) U^{(0)}(\mu_1) \left( U^{(0)}(\mu_2) \right)^{-1}
 J^{-1}(\mu_2) ,
\end{equation}
where
\begin{eqnarray}
U^{(0)}(\mu) &=& \exp \left(-\frac{\gamma^{0T}}{2 \beta_0} \ln \alpha_s(\mu) \right), \\
J(\mu) &=& \hat 1 + \frac{\alpha_s(\mu)}{4\pi} J^{(1)}
 + \left( \frac{\alpha_s(\mu)}{4\pi} \right)^2 J^{(2)} + \dots .
\end{eqnarray}
$U^{(0)}(\mu)$ is just the leading-order evolution operator, with the usual
ratio $\alpha_s(\mu_2)/\alpha_s(\mu_1)$ replaced by $1/\alpha_s(\mu)$.

We can then define
the following object appearing in the intermediate stages in our calculation:
\begin{equation}
  \hat M(\mu_i) = \left( U^{(0)}_{n_f}(\mu_i) \right)^{-1}
     \left(J_{n_f}(\mu_i)\right)^{-1} M_{n_f, n_f+1}(\mu_i)
     J_{n_f+1}(\mu_i) U^{(0)}_{n_{f+1}}(\mu_i) .
\end{equation}
Note that the matrix $\hat M(\mu_i)$ contains {\em all} the dependence
on the scale $\mu_i$ in the expression (\ref{eq:match}) for the Wilson
coefficient, and depends on no other scales.
As a consequence, it must by itself be independent of the renormalisation scale.
The way our calculation is set up, we obtain the matching matrix
with dependence on both $\alpha_s^{(n_f)}(\mu)$ and on $\alpha_s^{(n_f+1)}(\mu)$.
Expressing $\alpha_s^{(n_f+1)}$ in terms of $\alpha_s^{(n_f)}$
and taking into account both the explicit $\mu$-dependence and
the $\mu$-dependence of the quark mass appearing in the
matching, we observe that the scale dependence cancels, up to terms of
order ${\cal O}(\alpha_s^3)$.
The residual $\mu$-dependence, which is of order $\alpha_s^3$, gives
a complete estimate of higher-order effects at the threshold $\mu_i$.
We find that these residual scale dependences are at the percent level and that the value of $\varepsilon'/\varepsilon$ is only mildly shifted when going from NLO to NNLO. This
points to an excellent behaviour of the perturbation series not only
at the scale $\mu_b$, but also at the scale $\mu_c$.
This means at least for the dominant contribution
the perturbation theory seems to work. 

Our results are best expressed in terms of renormalisation-group and
scheme-independent Wilson coefficients and matrix elements,
\begin{eqnarray}
   \label{eq:chat}
   \hat C_i^{(n_f)} &=&
          \left( U_{n_f}^{(0)}(\mu) \right)^{-1}_{ik}
             \left(J_{n_f}(\mu)\right)^{-1}_{kj}  C_j^{(n_f)}(\mu) , \\
       \langle \pi \pi | \hat Q_i^{(n_f)} | K \rangle  &=&
            \langle \pi \pi | Q_j^{(n_f)}(\mu) | K \rangle \,
                \left( J_{n_f}(\mu) \right)_{jk} \left(
                  U^{(0)}_{n_{f}}(\mu) \right)_{ki} .
\end{eqnarray}
Note that we have not suppressed any $\mu$-dependence:
each hatted object is separately scale- and
scheme-independent\footnote{The explicit NNLO transformations of $J$, $C$ and $\langle Q \rangle$ under a change of scheme and the resulting scheme cancellation can be found in Ref.\cite{MH}.}. 
For this reason, relating the ``modern'' basis to
the ``traditional'' basis is trivial at the level of the hatted objects.
Perturbative uncertainties are fully contained in
the hatted coefficients, while non-perturbative uncertainties reside
in the hatted matrix elements. The two matching steps together with
RG evolution are then expressed as
\begin{equation}
   \hat C_i^{(3)} = \hat M^c_{ik} \hat M^b_{kj} \hat C^{(5)}_j .
\end{equation}
For dynamical charm, one simply removes the last matching step.

In fact, the hatted hadronic matrix elements could in principle
be obtained nonperturbatively on the lattice, obviating or reducing
the need to resort to technically demanding lattice-continuum
matching. If this is not done, the resultant extra ``perturbative''
uncertainty may be considered as part of the lattice error budget.

\section{$\varepsilon'/\varepsilon$ within the Standard Model}
Lattice computations of the relevant matrix elements for the analysis
of $\varepsilon'/\varepsilon$ are currently performed only in the 
isospin limit ($\alpha=0$, degenerate masses), and the Standard-Model prediction is based on the following expression
\begin{eqnarray}
\frac{\varepsilon'}{\varepsilon}=-i\frac{\omega_+}{\sqrt{2}|\varepsilon_K|}e^{i(\delta_2-\delta_0-\phi_{\varepsilon_K})}\left[\frac{\mathcal{\text{Im}}(\mathcal{A}_0)}{\mathcal{\text{Re}}(\mathcal{A}_0)}(1-\Omega_{\text{eff}})-\frac{1}{a}\frac{\mathcal{\text{Im}}(\mathcal{A}_2)}{\mathcal{\text{Re}}(\mathcal{A}_2)}\right] \, ,
\label{eq:CPVSM}
\end{eqnarray}
which is accurate up to ${\cal O}(1/\omega_+)$ corrections.
Here $\mathcal{A}_I\equiv \langle (\pi\pi)_I|\mathcal{H}_{\text{eff}}|K\rangle$
($I=0,2$) are the amplitudes for the two isospin states with their strong phases
$\delta_{0,2}$ removed. The latter, as well as
the phase $\phi_{\varepsilon_K}$ and magnitude
$|\varepsilon_K|$of $\varepsilon_K$ and the isospin ratio $\omega_+$
are all determined from experimental data.
The coefficients $a$ and $\Omega_{\text{eff}}$ are a (partial) parameterization
of corrections to the isospin limit, with values computed in chiral
perturbation theory in
\cite{Cirigliano:2003nn,Cirigliano:2003gt,Cirigliano:2011ny}.

The evaluation of the ratios $\mathcal{\text{Im}}({\cal
  A}_I)/\mathcal{\text{Re}}({\cal A}_I)$ is the central ingredient to the theory prediction of $\varepsilon'/\varepsilon$ and the formalism used for their determination is explained in reference \cite{Analysis}.
The quantities 
\begin{eqnarray}
\mathcal{\text{Re}}(\mathcal{A}_0) \approx \frac{4G_F}{\sqrt{2}} V_{ud} V_{us}^* (z_+\langle Q_+ \rangle_0 + z_-\langle Q_- \rangle_0),~~
\mathcal{\text{Re}}(\mathcal{A}_2) \approx \frac{4G_F}{\sqrt{2}} V_{ud} V_{us}^* z_+\langle Q_+ \rangle_2 
\label{eq:ReSM}
\end{eqnarray}
are completely dominated by the two operators $Q^u_{1,2}$.
($Q_{\pm} = \frac{1}{2}(Q_2^u \pm Q_1^u)$.) Operator identities
among the $V-A \times V-A$ operators imply that
the ratios
\begin{eqnarray}
\left(\frac{\mathcal{\text{Im}}(\mathcal{A}_2)}{\mathcal{\text{Re}}(\mathcal{A}_2)}\right)_{V-A}&=&\mathcal{\text{Im}}(\tau) \frac{3(y_9+y_{10})}{2z_+} \label{eq:Im0m} \\ 
\left(\frac{\mathcal{\text{Im}}(\mathcal{A}_0)}{\mathcal{\text{Re}}(\mathcal{A}_0)}\right)_{V-A}&=&\mathcal{\text{Im}}(\tau) \frac{2y_4}{(1+q)z_-}+\mathcal{O}(p_3) \label{eq:A0VmA} ,
\end{eqnarray}
where in the numerators only the terms involving the $V-A \times V-A$ operators
are kept, are almost free from hadronic uncertainties.
The factor $q$ in Eq.(\ref{eq:A0VmA}) is given in terms of the
current-current Wilson coefficients and operators,
$q\equiv (z_+(\mu)\langle Q_+(\mu)\rangle_0)/(z_-(\mu)\langle Q_-(\mu)\rangle_0)$. The small ratio $z_+/z_-$ implies that only a modest
accuracy is needed on the hadronic matrix elements entering the isospin-0
ratio through $q$. The term $\mathcal{O}(p_3)$ also involves a small Wilson
coefficient, which moreover multiplies a colour-suppressed hadronic matrix
element.

Conversely, the contributions coming from the $(V-A)\times(V+A)$
operators are very sensitive to long-distance effects (hadronic matrix
elements). To minimize the nonperturbative
uncertainties, one notes that ${\rm Re}({\cal A}_{0,2})$ govern CP-conserving
$K\to\pi \pi$ decays and are well determined by data.  One then has
\begin{eqnarray}
  \left(\frac{\mathcal{\text{Im}}(\mathcal{A}_0)}{\mathcal{\text{Re}}(\mathcal{A}_0)}\right)_{V+A}&=&
  -\frac{4 G_F}{\sqrt{2}}V_{ud} V_{us}^*\, \mathcal{\text{Im}}(\tau) \,y_6 \frac{\langle Q_6\rangle_0}{\mathcal{\text{Re}}(\mathcal{A}_0)}+\mathcal{O}(p_5)\\
  \left(\frac{\mathcal{\text{Im}}(\mathcal{A}_2)}{\mathcal{\text{Re}}(\mathcal{A}_2)}\right)_{V+A}&=&
  -\frac{4 G_F}{\sqrt{2}}V_{ud} V_{us}^*\, \mathcal{\text{Im}}(\tau) \, y_8^{\text{eff}} \frac{\langle Q_8\rangle_2}{\mathcal{\text{Re}}(\mathcal{A}_2)}.
\label{eq:V+A} 
\end{eqnarray}
Again, the omitted terms are suppressed by small Wilson coefficients
and/or colour-suppressed operator matrix elements. As a result, the
prediction for $\varepsilon'$ involves predominantly two hadronic matrix elements
(often parameterised in terms of parameters $B_6^{(1/2)}$ and $B_8^{(3/2)}$),
as well as perturbative Wilson coefficients $z_{1,2}$ and $y_{6,8}$. Our new
calculation essentiall removes the perturbative uncertainty on $y_6$. The
uncertainties on $z_{1,2}$ are already tiny, leaving $y_8$ and an improved
treatment of isospin-breaking corrections as the main objectives for the future.

Note that using ${\rm Re}({\cal A}_{0,2})$ from data also in the $V-A \times V-A$
operators introduces dependence on (mainly) the matrix element of the operator
$Q_4$, and its Wilson coefficient, and should be avoided. This is the main
reason why the prediction of \cite{Analysis} is more accurate than that of
RBC and UKQCD, and leads to a more pronounced tension with the data, in spite
of employing the same nonperturbative matrix elements and including
an error estimate for isospin-breaking corrections.

\section{Summary and outlook}
The experimental data for $\varepsilon'/\varepsilon$ is in mild tension
with the SM. This conclusion is possible because of recent progress in lattice
QCD, which has reduced the (still dominant) uncertainties on
long-distance non-perturbative contributions. With further progress
expected in the near future, currently subleading uncertainties, most
importantly higher-order perturbative short-distance contributions,
will start to dominate. We have described a calculation at NNLO
accuracy of the short-distance contributions to the QCD penguin
amplitudes, sufficient to achieve NNLO accuracy on
the isospin-zero $K \to \pi \pi$ decay amplitude
ratio. Our results leave the central value of the theoretical prediction
nearly unchanged, while greatly reducing the perturbative uncertainty on this
piece, leaving unknown NNLO corrections to the $I=2$
amplitude ratio to dominate the perturbative uncertainty on
$\varepsilon'$, at about half the previous NLO error.

Several further steps will be needed before the theoretical
uncertainty matches the experimental one, and perhaps a discrepancy
with the SM can be confirmed. One of them is the computation of the
$I=2$ amplitude ratio to NNLO accuracy. Another is a better treatment
of isospin-breaking corrections, and electromagnetic corrections in
particular. The latter affect both the Wilson coefficients and the
hadronic matrix elements, and require both perturbative and lattice
calculations, as well as a consistent interfacing of the two.

\ack
MC, SJ and AK would like to thank the organisers of KAON 2016 conference for this wonderful experience. 
The work of MC, MG and AK has been supported by the STFC consolidated
grant ST/L000431/1. SJ is supported by STFC consolidated grant
ST/L000504/1, a Weizmann Institute ``Weizmann-UK Making Connections''
grant, and an IPPP Associateship. 
We would like to thank Teppei Kitahara for notifying us of a typo in Eq.~\eqref{eq:2} and sharing their revised version.

\section*{Comment on the solution of the renormalisation group evolution}
\label{sec:comment-rge-its}

At this conference one of the authors of Ref. \cite{Kitahara:2016nld} presented a solution to the renormalisation group equation
\begin{equation}
  \label{eq:rge}
  \mu \, \frac{d}{d \mu} C = \gamma^T \, C
\end{equation}
for the Wilson coefficients $C$. Here 
\begin{equation}
  \label{eq:gammadef}
  \gamma = \sum_{m+n \ge 1} \gamma^{(m,n)} \left( \frac{\alpha_s}{4 \pi} \right)^m \left( \frac{\alpha_e}{4 \pi} \right)^n
\end{equation}
represent the anomalous dimensions relevant for $\varepsilon'/\varepsilon$.
While the presented solution incorporates some corrections of $\mathcal{O}(\alpha_e^2/\alpha_s^2)$ not all effects of this order have been take into account in this work. 
In the following we will give the correct solution for the evolution matrix \eqref{eq:uevolve} in the case of combined QCD and QED corrections, where 
\begin{equation}
  \label{eq:kansatz}
  \begin{split}
  J(\mu) =& \left[ 1 + \frac{\alpha_e}{4\pi} \left(J^{(1,1)}_{0} + J^{(1,1)}_{1} \ln \alpha_s + J^{(1,1)}_{2} (\ln \alpha_s)^2 \right) \right] 
            \left[ 1 + \frac{\alpha_s}{4\pi} \left(J^{(1,0)}_{0} + J^{(1,0)}_{1} \ln \alpha_s \right) \right] \times \\
          & \left[ 1 + \frac{\alpha_e}{\alpha_s} \left(J^{(0,1)}_{0} + J^{(0,1)}_{1} \ln \alpha_s \right) 
                     + \frac{\alpha_e^2}{\alpha_s^2} \left(J^{(0,2)}_{0} + J^{(0,2)}_{1} \ln \alpha_s \right) \right] \, 
  \end{split}
\end{equation}
follows the form chosen in Ref.~\cite{Kitahara:2016nld}. 
To determine a differential equation for $J(\mu)$ we use the scale invariance $\mu (d/d\mu) \hat{C} = 0$ of the Wilson coefficients defined in \eqref{eq:chat} and trade the scale dependence in the total derivative for an $\alpha_s$ dependence as 
\begin{equation}
  \label{eq:muderivative}
  \mu \frac{d}{d \mu} 
= - 2 \alpha_s^2 \frac{\beta_s}{4 \pi} \left( \frac{\partial}{\partial \alpha_s} - \frac{\alpha_e^2 \beta_e}{\alpha_s^2 \beta_s} \frac{\partial}{\partial \alpha_e} \right)
\, .
\end{equation}
Here the implicit $\mu$ dependence in $\alpha_e$ is now understood as $\alpha_e(\mu) = \alpha_e(\alpha_s(\mu))$ and $\beta_s$ and $\beta_e$ are defined via the  QCD $\mu (d/d \mu) \alpha_s = - 2 \alpha_s^2 \beta_s/(4 \pi)$ and QED $\mu (d/d \mu) \alpha_e = + 2 \alpha_e^2 \beta_e/(4 \pi)$ beta functions. 
The resulting equation for $J$ then reads:
\begin{equation}
  \label{eq:kequation}
  \frac{\partial J}{\partial \alpha_s} = \frac{\alpha_e^2 \beta_e}{\alpha_s^2 \beta_s} \frac{\partial J}{\partial \alpha_e} -  \frac{4\pi \, \gamma^T J}{2 \alpha_s^2 \beta_s} + \frac{J \gamma^{(1,0)^T}}{2 \alpha_s \beta_s^{(0,0)}} \, 
\end{equation}
and which implies a set of simple algebraic equations for the matrices $J^{(i,j)}_l$ which read
\begin{equation}
  \label{eq:2}
  \begin{split}
   0 &= \frac{\left[J_1^{\text{(1,0)}},\gamma^{(1,0)^T}\right]}{2 \beta_s^{(0,0)}}-J_1^{\text{(1,0)}} \\
   0 &= \frac{\left[J_0^{\text{(1,0)}},\gamma^{(1,0)^T}\right]}{2 \beta_s^{(0,0)}}+\frac{\beta_s^{(1,0)} \gamma^{(1,0)^T}}{2 \beta_s^{(0,0)^2}}-\frac{\gamma^{(2,0)^T}}{2 \beta_s^{(0,0)}}-J_0^{\text{(1,0)}}-J_1^{\text{(1,0)}} \\
   0 &= \frac{\left[J_1^{\text{(0,1)}},\gamma^{(1,0)^T}\right]}{2
   \beta_s^{(0,0)}}+J_1^{\text{(0,1)}} \\
   0 &= \frac{\left[J_0^{\text{(0,1)}},\gamma^{(1,0)^T}\right]}{2
   \beta_s^{(0,0)}}-\frac{\gamma^{(0,1)^T}}{2 \beta_s^{(0,0)}}+J_0^{\text{(0,1)}}-J_1^{\text{(0,1)}} \\
   0 &= \frac{\left[J_2^{\text{(1,1)}},\gamma^{(1,0)^T}\right]}{2
   \beta_s^{(0,0)}} \\
   0 &= \frac{\left[J_1^{\text{(1,0)}},\gamma^{(0,1)^T}\right]}{2
   \beta_s^{(0,0)}}+\frac{\left[J_1^{\text{(1,1)}},\gamma^{(1,0)^T}\right]}{2 \beta_s^{(0,0)}}-2 J_2^{\text{(1,1)}} \\
   0 &= \frac{\left[J_0^{\text{(1,0)}},\gamma^{(0,1)^T}\right]}{2
   \beta_s^{(0,0)}}+\frac{\left[J_0^{\text{(1,1)}},\gamma^{(1,0)^T}\right]}{2 \beta_s^{(0,0)}}+\frac{\beta_s^{(1,0)} \gamma^{(0,1)^T}}{2 \beta_s^{(0,0)^2}}+\frac{\beta _s^{\text{(0,1)}} \gamma ^{\text{(1,0)}^T}}{2 \beta_s^{(1,0)^2}}-\frac{\gamma^{(1,1)^T}}{2 \beta_s^{(0,0)}}-J_1^{\text{(1,1)}} \\
  0 &= \frac{\left[J_1^{\text{(0,2)}},\gamma^{(1,0)^T}\right]}{2
   \beta_s^{(0,0)}}-\frac{\gamma^{(0,1)^T} J_1^{\text{(0,1)}}}{2
   \beta_s^{(0,0)}}+\frac{J_1^{\text{(0,1)}} \beta_e^{(0,0)}}{\beta_s^{(0,0)}}+2
   J_1^{\text{(0,2)}} \\
  0 &= \frac{\left[J_0^{\text{(0,2)}},\gamma^{(1,0)^T}\right]}{2
   \beta_s^{(0,0)}}-\frac{\gamma^{(0,1)^T} J_0^{\text{(0,1)}}}{2
   \beta_s^{(0,0)}}+\frac{J_0^{\text{(0,1)}} \beta_e^{(0,0)}}{\beta_s^{(0,0)}}+ 2 J_0^{\text{(0,2)}}-J_1^{\text{(0,2)}} \, .
  \end{split} 
\end{equation}
Here the expansion coefficients $\beta_s^{(n,m)}$ and $\beta_e^{(n,m)}$ of the beta functions are defined via
\begin{equation}
  \label{eq:beta}
  \beta_s = \sum_{n,m \geq 0} \beta_s^{(m,n)} \left( \frac{\alpha_s}{4 \pi} \right)^m \left( \frac{\alpha_e}{4 \pi} \right)^n \quad \text{and} \quad 
\beta_e = \sum_{n,m \geq 0} \beta_e^{(m,n)} \left( \frac{\alpha_e}{4 \pi} \right)^m \left( \frac{\alpha_s}{4 \pi} \right)^n \,.
\end{equation}
\\[-5pt]
In the original version of Ref.\cite{Kitahara:2016nld} terms proportional to $\beta_e^{(0,0)}$ and $\beta_s^{(0,1)}$ were absent.
Our results agree with their revised version.

\end{document}